\documentclass{ifacconf}
\usepackage{graphicx}      
\usepackage{natbib}        

\usepackage{amsmath} 
\usepackage{amsfonts} 
\usepackage{algorithm}
\usepackage{algpseudocode}
\usepackage{subfig} 

\usepackage{scalefnt}
\usepackage{tikzscale}
\usepackage{tikz}
\usepackage{pgfplots}
\usepackage{changepage}
\usetikzlibrary{shapes.geometric}
\usetikzlibrary{calc}
\usetikzlibrary{positioning}
\usetikzlibrary{intersections}

\usepackage{url}

\usepackage{makecell}

\usepackage{fancyhdr}

\DeclareMathOperator{\diag}{diag}
\DeclareMathOperator{\argmax}{argmax}
\newcommand{\PCH}{\mathcal{P}}
\fancypagestyle{firstpage}
{
		\fancyhead[L]{}    
		\fancyhead[R]{© 2024. This work has been accepted to IFAC for publication under a Creative Commons Licence CC-BY-NC-ND}
	}
\begin{document}
\begin{frontmatter}

\title{Real-time MPC with Control Barrier Functions for Autonomous Driving using Safety Enhanced Collocation \thanksref{footnoteinfo}} 


\thanks[footnoteinfo]{This project has received funding from the European Union’s Horizon 2020 research and innovation programme under the Marie Skłodowska-Curie grant agreement ELO-X No 953348 and the Flemish Agency for Innovation and Entrepreneurship (VLAIO) under research project No. HBC.2021.0939 (BECAREFUL).}

\author[1,2]{Jean Pierre Allamaa} 
\author[2]{Panagiotis Patrinos} 
\author[3]{Toshiyuki Ohtsuka}
\author[1]{Tong Duy Son}



\address[1]{Siemens Digital Industries Software,  3001, Leuven, Belgium (e-mail: \{jean.pierre.allamaa, son.tong\}@siemens.com)}
\address[2]{Dept. Electr. Eng. (ESAT) - STADIUS research group, KU Leuven, 3001 Leuven, Belgium (e-mail: panos.patrinos@esat.kuleuven.be)}
\address[3]{Department of Informatics, Kyoto University, Kyoto, Japan (e-mail: ohtsuka@i.kyoto-u.ac.jp)}

\begin{abstract}                
The autonomous driving industry is continuously dealing with safety-critical scenarios, and nonlinear model predictive control (NMPC) is a powerful control strategy for handling such situations. However, standard safety constraints are not scalable and require a long NMPC horizon. Moreover, the adoption of NMPC in the automotive industry is limited by the heavy computation of numerical optimization routines. To address those issues, this paper presents a real-time capable NMPC for automated driving in urban environments, using control barrier functions (CBFs). Furthermore, the designed NMPC is based on a novel collocation transcription approach, named RESAFE/COL, that allows to reduce the number of optimization variables while still guaranteeing the continuous time (nonlinear) inequality constraints satisfaction, through regional convex hull approximation. RESAFE/COL is proven to be 5 times faster than multiple shooting and more tractable for embedded hardware without a decrease in the performance, nor accuracy and safety of the numerical solution. We validate our NMPC-CBF with RESAFE/COL on digital twins of the vehicle and the urban environment and show the safe controller's ability to improve crash avoidance by 91\%. Supplementary visual material can be found at~\url{https://youtu.be/_EnbfYwljp4}.

\end{abstract}


\end{frontmatter}
\thispagestyle{firstpage}
\pagenumbering{gobble}
\section{Introduction}
Advancements in autonomous driving (AD) and automated vehicles (AV) have been weakened by the acceptability of such intelligent components by human beings. This is mainly due to the safety-linked nature of AVs that are required to navigate with optimal performance, even in safety-critical situations such as urban driving in the presence of obstacles. Therefore, the AD motion planning module often needs to control the system at the limit of states and inputs. Nonlinear model predictive control (NMPC) is one solid control strategy allowing optimal and deterministic motion planning that optimizes over possibly conflicting stability, safety, and control objectives. There exists abundant work in the literature to tackle AD collision avoidance and NMPC has proven to be one promising approach, however, most of the work relies on simplified models, limited scenarios, and consider distance constraints with Euclidean norms. Although successful, the scalability of those approaches towards different applications, models (for e.g. shorter horizon), and embedded hardware deployment, is tight. In particular, the usage of a fast NMPC in the control loop is hindered by two main challenges: 1) lack of general methods to tackle urban driving with long planning horizons, and 2) the computation time of the underlying numerical optimization. 

One possible contained approach to combine a mid-level planner for collision-free trajectories, and a low-level control for trajectory tracking is known as control barrier functions (CBF). CBF approaches provide safety guarantees for the trajectories of the continuous-time nonlinear system beyond the predicted horizon in the NMPC through set invariance that exploits a Lyapunov-like condition for safety. CBFs have shown to be advantageous over other methods, when used within the MPC formulation (\cite{zeng_safety_critical_2021, son_safety_critical_2019}), or as a safety filter for end-to-end control approaches (\cite{cosner_end_end_2022}), yet not real-time capable. There are proven computationally efficient CBF formulations for collision avoidance as in~\cite{he_rule_based_2021}, but they do not scale well for urban driving and long planning horizons. Therefore, in this paper we present a real-time capable NMPC-CBF, for simultaneous path tracking and collision avoidance allowing to safely navigate through urban environments, while satisfying spatio-temporal input/state constraints.

In most MPC applications, the continuous-time model evolution equations are discretized using an integrator such as the Runge-Kutta $4^{th}$ order method (RK4). The original optimal control problem (OCP) can be transcribed into a nonlinear programming problem (NLP) using a popular approach known as Direct Multiple Shooting (DMS). For long predictions and dense time sampling, the discrete approaches result in a high number of optimization variables, making MPC challenging for real-time and embedded control, and tradeoffs are often made between the computational load and the closed-loop safety and performance. To circumvent the weakness of direct and discrete approaches, the collocation transcription method is used, allowing to reduce the number of optimization variables by posing the states and input trajectories as polynomials most famously with Pseudospectral collocation (PSC) that benefits from a fast convergence rate (\cite{Huntington2007AdvancementAA}). However, similar to DMS, PSC enforces safety constraints only at finite nodes. To tackle this issue, collocation based on B-splines with convex hulls have been used for collision avoidance as in~\cite{cichella_optimal_2021}. However, those approaches might lead to conservatism, and are 1) slow to converge in comparison with spectral methods and 2) not tractable for real-time constraint satisfaction due to the multidimensional Bernstein polynomials. 

Near the limits of performance where optimality is crucial, the safety of the predicted trajectory is not guaranteed with DMS and PSC. This can lead to uncomfortable and dangerous situations such as emergency braking or potentially, crash with other road users specifically in safety-critical conditions. Failure to generate safe control inputs limits the user's acceptability of such controllers, therefore this work presents a safe yet non-conservative MPC approach for autonomous driving. In this paper, we extend the collocation framework with safety envelope developed in~\cite{10178116}, to include general nonlinear constraints, and present a transcription framework for smooth MPC problems, called RESAFE/COL: Regional Envelope for SAFety Enhanced COLlocation. It allows solving for the original continuous-time OCP by generating safe trajectories of states and inputs as splines, satisfying the constraints over the complete prediction horizon through computationally efficient manipulation of the splines' coefficients. In addition, we introduce the concept of regional convex hulls to reduce the conservatism on the splines' extrema approximation, allowing dynamic solutions near the limits of handling towards a natural longitudinal and lateral driving. The contributions of this paper are three-fold: 
\begin{enumerate}
	\item Collocation framework with safety envelope over nonlinear constraints,
	\item Control barrier function development for safe autonomous driving in urban environment,
	\item Validation of RESAFE/COL with CBF in safety-critical situations, integrated in a real-time NMPC.
\end{enumerate}

This paper is organized as follows: in Sec.~\ref{sec:Background} we provide a background on a collocation method for NMPC and exponential control barrier functions. In Sec.~\ref{sec:resafe_col} we introduce the transcription method for real-time NMPC based on regional safety envelopes. We follow with Sec.~\ref{sec:autonomous_driving} in which we introduce the NMPC-CBF formulation for AD, and present the results in Sec.~\ref{sec:results}, before  concluding in Sec.~\ref{sec:conclusion}.

\section{Background}\label{sec:Background}
The navigation of the vehicle relies on a special set of splines that can guarantee a safe tracked trajectory in the look-ahead of the NMPC. The transcription into splines using collocation enhances the tractability of the NMPC problem for real-time applications, without decreasing the performance. In this section we first start by giving a background on the spectral orthogonal collocation with safety envelope, and briefly introduce control barrier functions. 
\subsection{Safety Envelope for Spectral Orthogonal Collocation}
We consider the nonlinear continuous Bolza problem, that optimizes a cost function $J$ over the states $x,u$ with a stage cost $l$ and terminal cost $\phi(x(t_f))$, to satisfy a set of (nonlinear) constraints $g, g_f$, by propagating the dynamics $f(x,u)$ over a time horizon $[0, t_f]$:
\begin{equation}
	\label{eq:generic_ocp}
	\left\{
	\begin{aligned}
		\min_{x(.),u(.)} &J(x,u) = \phi(x(t_f)) + \int_{t_0}^{t_f} l(x(t),u(t))dt \\
		\textrm{subject to } &\dot{x}(t) = f(x(t), u(t)),  \\
		&g(x(t), u(t)) \leq 0,\\
		&g_{f}(x(t_f)) \leq 0. \\
	\end{aligned}
	\right.
\end{equation}
We focus on the collocation approach to transcribe the OCP into an NLP as it does not require an embedded ODE solver. Instead, the problem is a dynamic optimization problem with polynomial representation of the state and control in each finite element. In particular, we enforce the control and states to be continuous on $[t_0, t_f]$, and transcribe the OCP into an NLP using spectral orthogonal collocation SOCSE as in~\cite{10178116}. For smooth problems, an accurate solution is obtained using a small number $N$ of collocation nodes, and the approach benefits from a spectral accuracy of convergence, faster than any power of $1/N$ (\cite{Huntington2007AdvancementAA}).

We define the orthogonal collocation scheme based on the truncated Legendre-series (TLS) of degree $M$ as in~\eqref{eq:legendre_spline} to approximate the solutions $x(\tau)$ and $u(\tau)$ of the continuous time OCP in a compact representation through the coefficients $\alpha$ rather than a discrete set of points:%
\begin{equation}%
	\label{eq:legendre_spline}
	x(\tau) = \sum_{k=0}^{M} \alpha_{k,x}\mathcal{L}_k(\tau) = \alpha_x^\top\mathbf{L}_Mv(\tau), (\tau\in[-1,1]),
\end{equation}%
and similarly for control inputs $u(\tau)$. The matrix $\mathbf{L}_M \in \mathbb{R}^{(M+1)\times(M+1)}$ is upper triangular, formed by the coefficients of $\mathcal{L}_k$ with respect to the normalized time horizon $\tau \in [-1, 1]$ of $t \in [t_0, t_f]$ in the OCP. In particular, the spanning basis are Legendre polynomials $\mathcal{L}_k$, which are orthogonal, satisfy $\mathcal{L}_k(1) = 1$ for any degree $k$, and in~\cite{Huntington2007AdvancementAA}, they are given by:%
\begin{equation}%
	\label{eq:legendre_polynomial}
	\mathcal{L}_k(\tau) = \frac{1}{2^k k!} \frac{d^k}{d\tau^k}[(\tau^2-1)^k].
\end{equation}%
An optimization over the polynomial with an orthogonal basis 1) enhances numerical stability and, 2) shrinks the higher order terms to zero if they are not significant on the output. Moreover, the TLS is parametrized by the coefficients $\alpha_x = \begin{bmatrix}\alpha_0 & \cdots & \alpha_M\end{bmatrix}^\top \in \mathbb{R}^{(M+1)\times N_x}$ , $\alpha_u \in \mathbb{R}^{(M+1)\times N_u}$, and $v(\tau) = \begin{bmatrix} 1 & \tau & \tau^2 & \cdots & \tau^M \end{bmatrix}^\top$ is a vector with a geometric progression of the normalized time instance $\tau = (2t/t_f -1)$ with $t_0=0$. In addition, $N_x, N_u$ are the number of state and control variables respectively and $M$ is the degree of the fitting polynomial. The dynamics equations of the OCP are satisfied by differentiating~\eqref{eq:legendre_spline} at the $N$ collocation points $\tau_i$ to satisfy the dynamics:
\begin{equation}
	\label{eq:dynamics_collocation}
	\dot{x}(\tau_i) = \alpha_x^\top\mathbf{L}_M\dot{v}(\tau_i) = \frac{t_f}{2}f(x(\tau_i), u(\tau_i)).
\end{equation}
Note that $dt/d\tau=t_f/2$ accounts for the timescale transformation. The $N$ collocation points $\tau_i$, are neither random, nor uniform, but are specifically set as the Legendre-Gauss-Lobatto (LGL) nodes, defining the spectral grid of a Lobatto Pseudospectral method (LPM). The collocation points $\tau_i, i=2,\dots,N-1$ are uniquely chosen to be the roots of $\dot{\mathcal{L}}_{N-1}$, the derivative of the Legendre polynomial $\mathcal{L}_{N-1}$ of degree $N-1$, c.f~\eqref{eq:legendre_polynomial}. The two remaining collocation points are the boundaries $[-1,1]$ such that $\mathrm{T} =: \begin{bmatrix} -1 & \textrm{Roots of } \dot{\mathcal{L}}_{N-1} & 1\end{bmatrix}$ are the collocation nodes (\cite{10178116}). LPM can achieve exact estimation up to machine precision, as long as the true polynomial is up to degree $2N-3$, with a relatively small number of nodes (\cite{Huntington2007AdvancementAA}) and a truncation error $O(h^{2N-2})$. Importantly, from~\cite{10178116} and with the particular choice of parametrization as in~\eqref{eq:legendre_spline}, the resulting TLS is contained in its convex envelope, having as its boundaries the maximum and minimum elements of its convex hull $\PCH_M \in \mathbb{R}^{(M+1)}$:
\begin{align}
	\label{eq:safety_envelope}
	&\min \{\PCH_M\} \leq x(\tau) \leq \max \{\PCH_M\}, \;\; \forall \tau \in [-1,1],\\ 
	&\mathrm{where } \;\; \PCH_M =  \mathcal{B}\mathcal{E}^\top\mathbf{L}_M^\top\alpha_x = \mathcal{C}_M\alpha_x.
\end{align}
As proven in~\cite{cargo_bernstein_1966}, the matrix $\mathcal{B} \in \mathbb{R}^{(M+1) \times (M+1)}$ is a mapping from the coefficient of a polynomial to the Bernstein extrema estimations $b_j$, such that for a polynomial $P(t)=a_0+a_1t+a_2t^2+\dots+a_Mt^M$ of degree $M\geq0$ with real coefficients $a_j, j=0,\hdots,M$, and defined over the time interval $[0,1]$:%
\begin{subequations}
	\label{eq:bernstein_coefficients}
	\begin{align}
		b_j &= \sum_{k=0}^{j}a_k \binom{j}{k}\bigg/\binom{M}{k}, \, j=0,1,\dots,M,\\
		b &= \mathcal{B}a, \, \mathcal{B} \in \mathbb{R}^{(M+1) \times (M+1)}.
	\end{align}
\end{subequations}
The lower triangular matrix $\mathcal{E} = \mathcal{E}^{-1:1} \in \mathbb{R}^{(M+1) \times (M+1)}$ is a time transformation for the vector $v(\tau)$ in~\eqref{eq:legendre_spline} allowing a conversion from any interval of normalized time instances $[\tau_k,\tau_{k+1}]\subset [-1,1]$ with $\tau = (2\tau_*-1) $ to $\tau_* \in [0, 1]$, and particularly for the case of the boundaries such that $\tau_k=-1, \tau_{k+1}=1$. Using the Binomial theorem, the entry on the $i^{th}$ row and $j^{th}$ column of the matrix $\mathcal{E}^{\tau_k:\tau_{k+1}}$ is:
\begin{subequations}
	\begin{align}
		\label{eq:binomial_time_transform_a}
		\mathcal{E}_{ij}^{\tau_k:\tau_{k+1}} =  \sum_{i=0}^{M}\sum_{j=0}^{i}\binom{i}{j}(\tau_{k+1}-\tau_k)^{j}(\tau_k)^{i-j},\\
		\label{eq:binomial_time_transform_b}
		\textrm{such that: }v(\tau) = \begin{bmatrix} 1 & \tau  & \cdots & \tau^M \end{bmatrix}^\top = \mathcal{E}^{\tau_k:\tau_{k+1}} v(\tau_*). 
	\end{align}
\end{subequations}
The extrema approximation is tight and exact, if the true maximum and minimum are at the boundary elements of $\PCH_M$, and that is when the TLS is monotonic within the time interval. 
Similarly, the bounds on the states' and inputs' derivatives can be calculated by rewriting $\mathbf{L}_M$ in~\eqref{eq:legendre_spline} and~\eqref{eq:safety_envelope} in terms of the coefficients of $\mathcal{L}^{(i)}_{k}(\tau)$ instead of $\mathcal{L}_{k}(\tau)$.  Similar matrices $\mathcal{C}_{M,i}$ can be calculated offline, allowing the efficient evaluation of the convex hull of the $i^{th}$ derivative within the NLP.
The previous result is exploited to transcribe the linear inequality constraints of the continuous time OCP into convex constraints on the spline coefficients $\alpha$. However, there remains a question on how to efficiently exploit $\alpha, \PCH_M,$ as to include general nonlinear constraints within the optimization routine, which will be elaborated in this paper.
\subsection{Exponential Control Barrier Function}
Motivated by Control Lyapunov Functions (CLF) for guaranteeing the stability of a system, CBFs are employed to guarantee the safety of a control action through forward invariance conditions. A CBF extension for non-affine nonlinear systems with relative degrees higher than 1 has been derived in the work of~\cite{son_safety_critical_2019}.  

Consider a nonlinear dynamic system $\dot{x} = f(x,u)$, a set $\mathcal{C} \in  \mathbb{R}^{n}$ is forward invariant, if for every $x(0) = x_0 \in \mathcal{C}$, the state trajectory $x(t) \in \mathcal{C}, \forall t$. Moreover, the continuously differentiable function $h(x): \mathbb{R}^{n} \rightarrow \mathbb{R}$ is a CBF iff: 
\begin{subequations}%
	\begin{align}
		\mathcal{C} &= \{x\in \mathbb{R}^{n} \rvert h(x)\geq 0 \},\\
		\partial\mathcal{C} &= \{ x \in \mathbb{R}^{n} \rvert h(x) = 0 \}, \\
		\mathrm{Int}(\mathcal{C}) &= \{x \in \mathbb{R}^{n} \rvert h(x) > 0\}.
	\end{align}
\end{subequations}%
To guarantee forward invariance of a safety constraint $h(x)$ within a finite-horizon MPC, \cite{son_safety_critical_2019} derive the condition for an exponential CBF:
\begin{equation}\label{eq:CBF_definition_cnstrnt}
	L_f^rh(x,u) + k_1h(x) + k_2L_fh(x) + \dots + k_{r-1}L_f^{r-1}h(x) \geq 0.
\end{equation}
The Lie derivative $L_f$, and high-order Lie derivatives $L_f^k$ of the barrier function $h(x)$ with respect to the vector function $f(x,u)$ are:
\begin{align}
	L_f h(x,u) &= \frac{\partial h(x)}{\partial x}f(x,u),\\
	L_f^kh(x,u) &= \frac{\partial L_f^{k-1}h(x,u)}{\partial x} f(x,u).	
\end{align}
with $L_f^0h(x,u) = h(x)$. Furthermore, $r$ is such that the constraint $h(x)$ has a relative degree $1\leq r \leq n$. In other words, the input does not explicitly appear in the first $r$ Lie derivatives and that is $L_f^ih(x,u) = L_f^ih(x)$ for $0\leq i \leq r-1$. Moreover, there exists a vector $[k_1, k_2, \dots, k_{r-1}]$ such that the safety condition $h(x)\geq0$ is satisfied. 


\begin{figure}
	\centering
	\subfloat[][DMS and PSC]{	\includegraphics[width=0.23\textwidth]{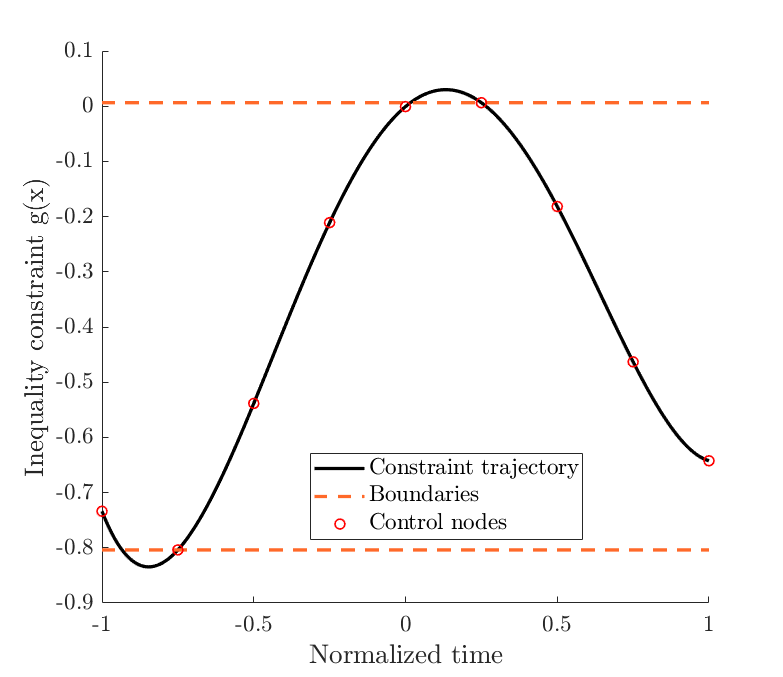}\label{fig:rndom_MS}}
	\subfloat[][RESAFE/COL (ours)]{	\includegraphics[width=0.23\textwidth]{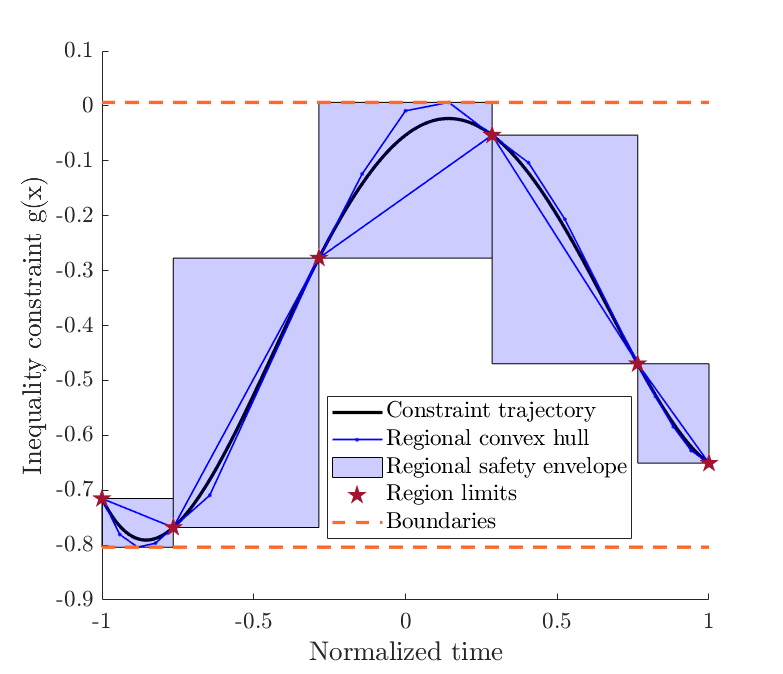}\label{fig:rndom_rc}}
	\caption{Continuous constraint satisfaction with RESAFE/COL with one finite element in comparison with DMS and PSC: DMS and PSC optimize over finite control nodes, and RESAFE/COL over the Legendre-series' coefficients. In Fig.~\ref{fig:rndom_MS}, DMS and PSC fail to capture the internode dynamics and violate the continuous-time constraint. In Fig.~\ref{fig:rndom_rc}, RESAFE/COL, through regional safety envelopes accurately estimates the (nonlinear) constraint's extrema, and returns a continuous-time feasible solution}
	\label{fig:resafecol_vs_ms_splines}
\end{figure}


\section{Regional envelope for nonlinear constraints}\label{sec:resafe_col}
In the standard direct approaches such as DMS or PSC, path inequality constraints are only enforced at the shooting or collocation points (\cite{Huntington2007AdvancementAA}). In some work such as~\cite{ruof_real_time_2023}, temporary numerical constraints violations are tolerated. However, for automotive industry standards or for other safety-critical robotic applications, this might not always be permissible. In this section, we present a transcription approach for NMPC, allowing to capture the extrema of the states, inputs, and nonlinear constraints in their convex hulls. Those extrema would serve a basis for a sped-up numerical optimization.

For non-smooth systems, and solutions that cannot be fit with a single TLS, the normalized NMPC horizon $[-1, 1]$ can be divided into smaller finite elements, to create a piecewise polynomial, namely a Legendre-spline. On each element, every state and input is parametrized by a TLS, the coefficients of which serve as optimization variables $\alpha$. Without loss of generality, we assume one finite element over the time horizon, and seek to minimize the approximation error on the spline's extrema.
Similar to Theorem 4 in~\cite{Rivlin1970BoundsOA}:
\begin{prop}\label{prop:decreasing_error_resafecol}%
	 The error on the regional extrema approximation for a truncated Legendre-series, decreases proportionally to $1/k^2$ for an increasing number of regions $k$ within one finite element on $[0, t_f]$.
\end{prop}%
Instead of estimating the spline's extrema on the full time horizon as in SOCSE, we propose a regional approximation defined by the regional convex hulls $\PCH_M^k$. The time interval $[-1, 1]$ is divided into $K$ regions as: $[t_0, t_1, \dots, t_{k+1}]$, that are not necessarily equidistant, and define a region $k$ to be a sub-interval $[t_k, t_{k+1}] \subset [-1, 1]$, with $t_{k+1} \ge t_k$. 
At the expense of computing additional mapping matrices $C_M^k$, but offline, we can compute the regional convex hulls as:
\begin{align}%
	\label{eq:regional_safety_envelope}
	 \PCH_M^k =  \mathcal{B}(\mathcal{E}^{t_k:t_{k+1}})^\top\mathbf{L}_M^\top\alpha = \mathcal{C}_M^k\alpha.
\end{align}%
From practice, a reasonable choice is to set the region bounds $t_k$ as the LGL nodes of the Legendre polynomial of order $K$, where $K$ is a hyperparameter for RESAFE/COL. The regional extrema approximation using a linear mapping over $\alpha$ are shown in Figure~\ref{fig:resafecol_vs_ms_splines}, and can be applied to state and input box constraints, as well as path and general nonlinear constraints.  The transfer to a regional safety envelope formulation has a three-fold benefit:
\begin{itemize}
	\item the conservatism on the spline extrema approximation is reduced,
	\item the open-loop trajectory, can be more dynamic with change of derivative, as large high-order coefficients lead to a conservative extrema approximation if only one region is used,
	\item regional convex hulls are computed using the original optimization variables $\alpha$, allowing for time varying constraints without the burden of additional optimization variables.
\end{itemize}

Therefore, RESAFE/COL is a generalization of SOCSE with more than one region, and allows the inclusion of nonlinear constraints as will follow.
   
\begin{thm}\label{thm:resafe_col}   
	Using the regional convex hull $\PCH^k$ of the state (or input) spline, the nonlinear constraint $g(x)$ is included in its continuous safety envelope defined by:
	\begin{align}\label{eq:resafecol_bounded_constraint}
		\min\{g(\PCH^k)\} - &\frac{d_k^2}{2}\max_{x\in \mathcal{X}^k}\left| \frac{d^2g}{dx^2} \right| \\ \nonumber
		&\leq g(x)\leq \\
		&\max\{g(\PCH^k)\} + \frac{d_k^2}{2}\max_{x\in \mathcal{X}^k}\left| \frac{d^2g}{dx^2} \right|, \nonumber
	\end{align}
where $\{\mathcal{X}^k: x(t)\lvert x(t_k) \leq x(t) \leq x(t_{k+1})\}$ and $d_k = \max\{\bar{\PCH}^k_{i+1} - \bar{\PCH}^k_{i} \}$, where $\bar{\PCH}^k_{i+1}$ is a sorted $\PCH^k$ with $i=[1,\dots,M]$.
\end{thm}

\begin{pf}    
	Let $\xi = \argmax(g(x))$ for $x \in \mathcal{X}^k$. Given the definition of $d_k$, it stands that $d_k \geq \lvert \PCH_i^k-\xi \lvert$.
	
	Through the generalized $2^{nd}$ order Taylor series expansion:
	\begin{equation}
		g(\PCH^k_i) = g(\xi) + g'(\xi)(\PCH^k_i - \xi) + \frac{g''(\beta)}{2}(\PCH^k_i - \xi)^2, \, \beta \in \mathcal{X}^k.
	\end{equation}
	
	\begin{itemize}
	\item If $g'(x) \neq 0$ for $x \in \mathcal{X}^k$, then the extrema are at the boundaries and the extrema's approximation is tight and that is $\max g(x) = \max \{g(\PCH^k)\}$, which verifies the upper bound in~\eqref{eq:resafecol_bounded_constraint} and the extrema is an element of the convex hull.
	\item If there is a change of curvature, $g'(x) = 0$ for $x \in \mathcal{X}^k$, with $g'(\xi) = 0$, the extrema is contained in an enlarged convex hull as: 
	\begin{align}
		g(\xi) = g(\PCH^k_i) - &\frac{g''(\beta)}{2}(\PCH^k_i - \xi)^2 \\ &\leq \max\{g(\PCH^k)\} + \frac{d_k^2}{2}\cdot \max_{x\in \mathcal{X}^k} \left| g''(x) \right|.\nonumber
	\end{align}
	\end{itemize}
	The lower bound of~\eqref{eq:resafecol_bounded_constraint} is proven similarly. \qed
\end{pf}
Note that for clarity, Theorem~\ref{thm:resafe_col} was proven for the scalar case with $g(x):\mathbb{R}\rightarrow \mathbb{R} $, however, it extends to multidimensional systems by replacing $\max_{x\in \mathcal{X}^k}\lvert \frac{d^2g}{dx^2}\lvert$ with the maximum eigenvalue of the Hessian of the nonlinear constraint $g(x, u)$. Moreover, as the number of regions increases, $d_k$ decreases and converges to zero, thus the approximation with the regional convex hull gets tighter.



\section{Application of RESAFE/COL to Autonomous Driving}\label{sec:autonomous_driving}
In this section, we apply the derived transcription approach to the case of automated driving in an urban environment in the presence of road users. We first start by introducing the vehicle model in the NMPC prediction, then follow with the collision avoidance protocol and present the resulting NMPC formulation and the NLP.
\subsection{Vehicle model}
We represent the car dynamics in the NMPC by a 3 DoF dynamic single-track model, in the curvilinear error frame:%
\begin{align}
		\label{eq:BicycleModelStates}
		M\Dot{v}_x &= (F_{xf} \cos\delta + F_{xr} -F_{yf} \sin\delta - F_{res} + Mr v_y), \nonumber\\
		M\Dot{v}_y &= (F_{xf} \sin\delta + F_{yr} + F_{yf} \cos\delta - Mr v_x), \\\nonumber
		I_z\Dot{r} &= (L_f (F_{yf} \cos\delta + F_{xf} \sin\delta) - L_r F_{yr}), \\\nonumber
		\Dot{s} &= (v_x \cos\theta -v_y \sin\theta)/(1-\kappa_c w), \\\nonumber
		\Dot{w} &= v_x \sin\theta + v_y \cos\theta, \\\nonumber
		\Dot{\theta} &= r - \Dot{\psi_c} = r - \kappa_c \Dot{s}.\nonumber
\end{align}%
\begin{figure}
	\begin{center}
		\includegraphics[width=6.4cm,trim={0cm 0cm 0cm 0cm},clip]{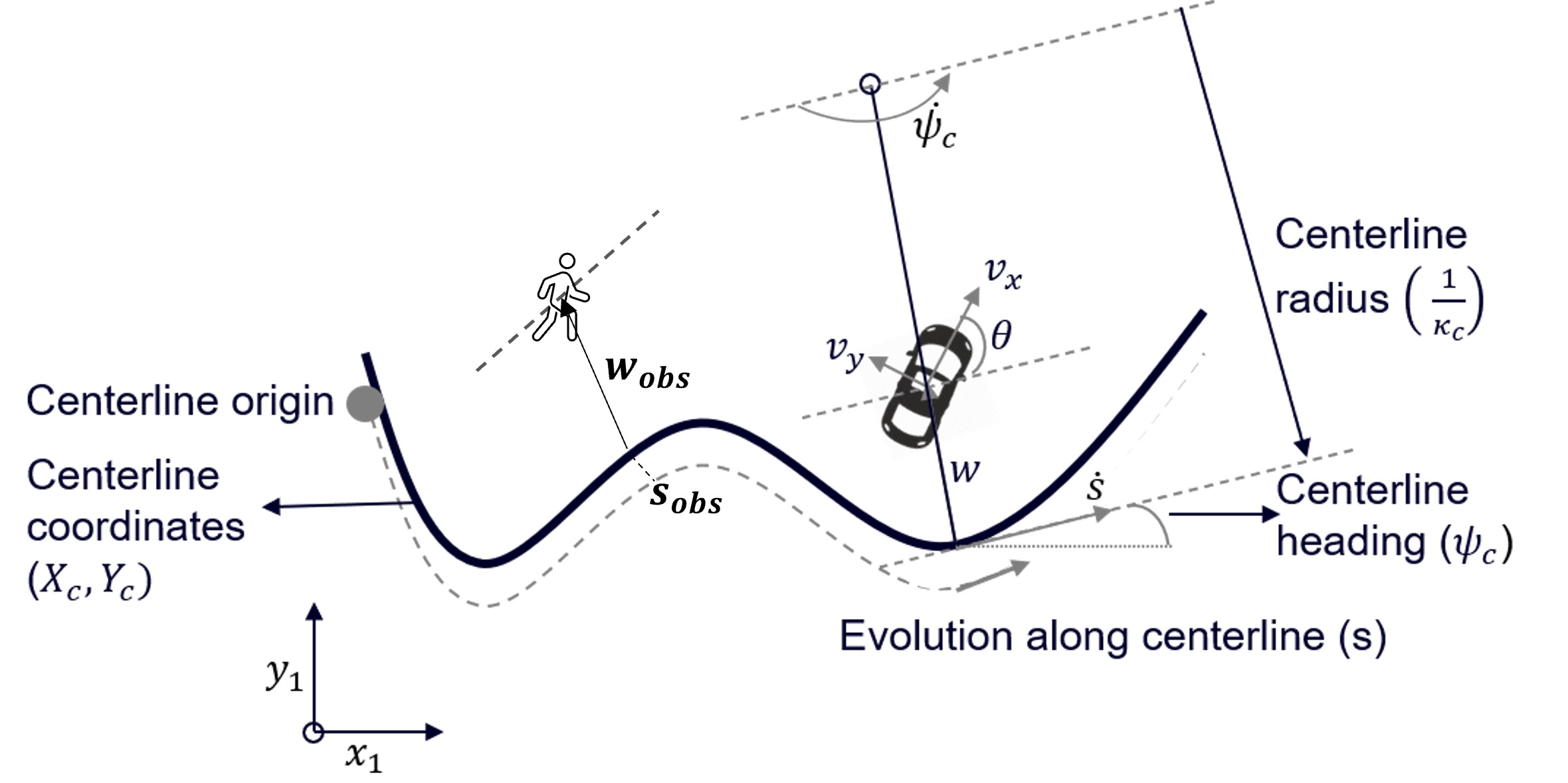}    
		\caption{Transformation to Curvilinear frame} 
		\label{fig:curvilinear_frame}
	\end{center}
\end{figure}

The first three equations of~\eqref{eq:BicycleModelStates} dictate the dynamics (velocities $v_x, v_y$ and yaw rate $r$) in the car body frame, and the last three represent the vehicle's kinematics in the curvilinear frame as presented in Figure~\ref{fig:curvilinear_frame}. Moreover, $M$ is the vehicle's mass, $I_z$ is the inertia about the z-axis, and $L_r,L_f$ are the distances from the center of gravity to the rear and front axles. 
For control purposes, we formulate the kinematics in the curvilinear or Frenet frame, that is equivalent to an error frame with respect to the desired path. That is, $s$ tracks the evolution along the path, $w$ and $\theta$ are the distance and heading deviation from the path respectively. This transformation allows an easier navigation on curved roads as it is only parametrized by the road curvature $\kappa_c(s)$ and exposes convex constraints over the path deviation as a tube $w_l\leq w \leq w_r$ where $w_l$ and $w_r$ are the left and right track limits. Additionally, it allows for spatial prediction in the NMPC horizon.  For smoother driving style, the NMPC dynamics model is augmented with the input rates  $u =[\dot{t}_r,\Dot{\delta}]$, the derivatives of the steering angle control $\delta$ and the normalized acceleration $t_r$ over the longitudinal forces $F_{x,\{r,f\}}$. The lateral forces $F_y$ are assumed linear with respect to the wheel slip angle.
As explained in~\cite{10178116}, we fuse the dynamic model in~\eqref{eq:BicycleModelStates} with the kinematic one to allow the model to be numerically stable near zero speeds. Hence, the single-track curvilinear dynamics between the state vector $x = [v_x, v_y, r, s, w, \theta, \delta, t_r]^\top$ and input $u$ are:%
\begin{equation}%
	\label{eq:dynamics}
	\Dot{x} = f(x,u) = \lambda f_{dyn}(x,u) + (1-\lambda)f_{kin}(x,u).
\end{equation}%
The control objective of the NMPC is a path following navigation with zero reference deviation from the path with velocity tracking as a non-zero term in the reference $x^r$, considering energy efficiency in terms of inputs. Therefore, we set the quadratic stage cost with weighting matrices $Q\in  \mathbb{R}^{8\times8}\succeq 0,R \in\mathbb{R}^{2\times2}\succ 0$ as:%
{\begin{equation}\label{eq:stage_cost_nmpc}%
		l(x(\tau),u(\tau)) = \lVert(x(\tau) - x^r(\tau))\lVert^2_Q 
		+ \lVert u(\tau)\lVert^2_R. 
\end{equation}}%
\subsection{CBF for collision avoidance}
In order to combine the safety constraint with the NMPC's state prediction, and input and state constraints, we opt to incorporate the CBF within the NMPC framework. We consider collision avoidance for AD, in an environment with multiple static and dynamic agents. In the work of~\cite{reiter_frenet_cartesian_2023}, ellipsoidal safety constraints have shown be superior to other geometric shapes. For this reason, we augment obstacles through ellipsoidal boundaries, centered at $(s_{obs}, w_{obs})$, with axes lengths $a$ and $b$. We set the safety constraint in the Curvilinear frame, by projecting the obstacles' Cartesian position in the global frame on the path frame driven by the autonomous vehicle. That is, $s_{obs}$ and $w_{obs}$ are the obstacle's  evolution along the path and its deviation from the closest point on the path respectively. The safety constraint is formulated as:
\begin{equation}%
	\label{eq:barrier_constraint}
	h(x) = \frac{(s-s_{obs})^2}{a^2} + \frac{(w-w_{obs})^2}{b^2} - 1 \geq 0.
\end{equation}%
Finally, the exponential CBF constraint is:
\begin{align}%
	\label{eq:cbf_constraint}
	h_{CBF}(x,u) &= L_f^2h(x,u) + k_1L_fh(x) + k_2h(x) \geq 0,\\
	L_fh(x) &=  \frac{2(s-s_{obs})\dot{s}}{a^2} + \frac{2(w-w_{obs})\dot{w}}{b^2},\\
	L_f^2h(x,u) &= \frac{2(s-s_{obs})\ddot{s} + 2\dot{s}}{a^2} + \frac{2(w-w_{obs})\ddot{w} + 2\dot{w}}{b^2}.
\end{align}%
In~\cite{son_safety_critical_2019}, a change of variable was required to get rid of the second order derivatives. In comparison, and given that each state trajectory is a TLS defined by~\eqref{eq:legendre_spline}, we can efficiently calculate $\dot{s}, \dot{w}, \ddot{s}, \ddot{w}$ by differentiating the Vandermonde vector $v(\tau)$ without relying on the dynamics. That is, the $k^{th}$ derivative of the spline, can be written in a compact form as $d^k x/dt^k = \frac{2}{t_f}\cdot\alpha \mathcal{L} (d^kv(\tau)/d\tau^k)$ where $d^kv(\tau)/d\tau^k$ is trivially calculated given the geometric form of the vector $v(\tau)$.  In relation with~\eqref{thm:resafe_col}, the barrier function in~\eqref{eq:barrier_constraint} is multidimensional but quadratic, hence with a diagonal Hessian. Therefore, the maximum eigenvalue of the constraint's Hessian can be calculated as the maximum of $\{2/a^2, 2/b^2\}$.

\subsection{NMPC formulation}
We cast the continuous time OCP into an NLP using RESAFE/COL. That is, the integral of the Lagrange term $l(x(\tau),u(\tau))=:l(\tau)$ in the OCP cost function is given by the exactness of the Gauss quadrature rule as:%
\begin{equation}%
	\label{eq:gauss_quadrature}
	\int_{-1}^{1}l(\tau)d\tau = w_1 l(-1) + w_N l(1) + \sum_{i=2}^{N-1}w_il(\tau_i).
\end{equation}%
The dynamics are satisfied according to~\eqref{eq:dynamics_collocation}. Moreover, the inequality constraints on the optimization variables $\alpha_x, \alpha_u$, the coefficients of the TLS are set through the regional envelopes in~\eqref{eq:regional_safety_envelope} and~\eqref{eq:resafecol_bounded_constraint}.
This approach benefits from the offline computation of the mapping matrices $\mathcal{C}^k_M \in\mathbb{R}^{(M+1)\times (M+1)}, \forall k\in[1, \hdots,K]$, according to \eqref{eq:regional_safety_envelope}, given the spline degree $M$ and the number of regions $K$. The box constraints on the states and inputs are replaced by linear constraints over $\alpha_x, \alpha_u$, and the nonlinear constraints maintain their structure, but are applied over the regional convex hull of the trajectories $\PCH^k_{M,x}, \PCH^k_{M,u}$ and their derivatives $\PCH^k_{M,\dot{x}}, \PCH^k_{M,        \ddot{x}}$. In other words, each continuous time constraint is transcribed by replacing the variable with its convex hull. The resulting NLP is:%
\begin{equation}%
	\label{eq:resafecol_nlp}
	\left\{
	\begin{aligned}
		\min_{\alpha_x, \alpha_u} &\frac{t_f}{2}[\sum_{i=2}^{N-1}w_il(\tau_i)+ \phi(\alpha_x^\top\mathbf{L}_M v(1))\\
		&\qquad + w_1 l(x_0, \alpha_u^\top\mathbf{L}_M v(-1)) + w_N l(1)]\\
		\textrm{s.t. } &\alpha_x^\top\mathbf{L}_M v(-1) = \bar{x}_0,\\
		&\alpha_x^\top\mathbf{L}_M\dot{v}(\tau_i) = \frac{t_f}{2}f(\alpha_x^\top\mathbf{L}_M v(\tau_i), \alpha_u^\top\mathbf{L}_M v(\tau_i))\\ 
		&\underline{x} \leq (\mathcal{C}^k_M \alpha_x)^\top\leq \overline{x},\,  \;\forall k\in[1, \hdots,K],\\
		&\underline{u} \leq (\mathcal{C}^k_M \alpha_u)^\top\leq \overline{u},\, \;\forall k\in[1, \hdots,K],\\
		&h(\mathcal{C}^k_M \alpha_x) \geq 0 \quad \textrm{from}\, \eqref{eq:regional_safety_envelope}, \eqref{eq:barrier_constraint},\\
		&h_{CBF}(\mathcal{C}^k_M \alpha_x)  \geq 0 \quad \textrm{from} \, \eqref{eq:resafecol_bounded_constraint}, \eqref{eq:cbf_constraint},
	\end{aligned}\right.
\end{equation}%
where $l(\tau_i) = l(\alpha_x^\top\mathbf{L}_M v(\tau_i), \alpha_u^\top\mathbf{L}_M v(\tau_i)), \forall i \in [1,N]$, the evaluation of the cost function at a particular collocation point. It is important to note that the barrier constraint and the CBF constraints are applied to the convex hulls of the states and their derivatives, but the second order terms from~\eqref{eq:resafecol_bounded_constraint} are left out as slack variables. The NLP is solved with a Sequential Quadratic Programming (SQP) method using the OSQP QP solver (\cite{osqp}) as SQP is suited for embedded applications such a real-time control for AD.

We set the major-axes of the ellipses enlarging the static and dynamic obstacles to $a = 3\,\textrm{m}$, $b = 2\,\textrm{m}$, the coefficients of the CBF constraint $k_1 = 1.6, k_2 = 1.1$ and the weighting matrices of the cost function $Q = \diag(3.1,   10,   10,         0,    5.2,   48,    0.9,    1.5), R=\diag(1,    1)$. Moreover, the control action is executed every $50\,\textrm{ms}$. Finally, we distinguish between two situations based on the prediction horizon $t_f$: long look-ahead with $t_f=3\,\textrm{s}$ and short look-ahead for safety-critical situations with $t_f=1.75\,\textrm{s}$.

\section{Results and comparison with DMS and PSC}\label{sec:results}
We define four vehicles, each represented as a Digital Twin of an electrical SimRod vehicle. That is, each vehicle is a high-fidelity 15DoF simulator in Simcenter Amesim, and navigates in the traffic scenario simulator Simcenter Prescan. The four vehicles are defined as:
\begin{itemize}
	\item Vehicle 1, labeled RESAEF/COL+CBF: white vehicle (black plots), NMPC with the exponential CBF and position constraints~\eqref{eq:barrier_constraint},\eqref{eq:cbf_constraint}, transcribed using RESAFE/COL, with TLS of degree $M=5$.
	\item Vehicle 2, labeled PSC: blue vehicle (blue plots), NMPC with position constraint~\eqref{eq:barrier_constraint}, transcribed using Pseudospectral collocation, and polynomials of order 5.
	\item Vehicle 3, labeled RESAFE/COL: red vehicle (red plots), NMPC with position constraint~\eqref{eq:barrier_constraint}, transcribed using RESAFE/COL, and TLS of degree 5.
	\item Vehicle 4, labeled DMS: green vehicle (green plots), NMPC with position constraint~\eqref{eq:barrier_constraint}, transcribed using Direct Multiple Shooting.
\end{itemize}

\begin{figure}
	\centering
	\subfloat[][Long look-ahead]{	\includegraphics[width=0.23\textwidth]{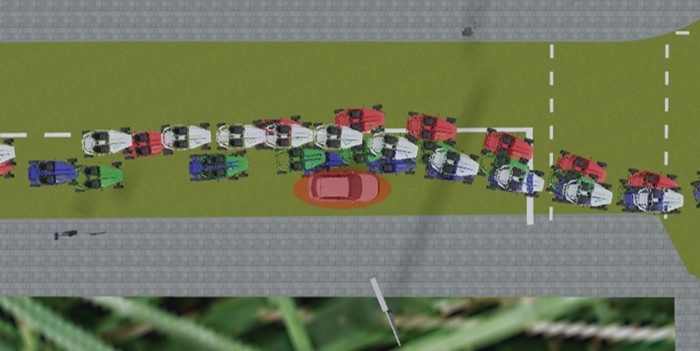}\label{fig:cbf_vs_pos_constraint_munich_a}}
	\subfloat[][Short look-ahead]{	\includegraphics[width=0.23\textwidth]{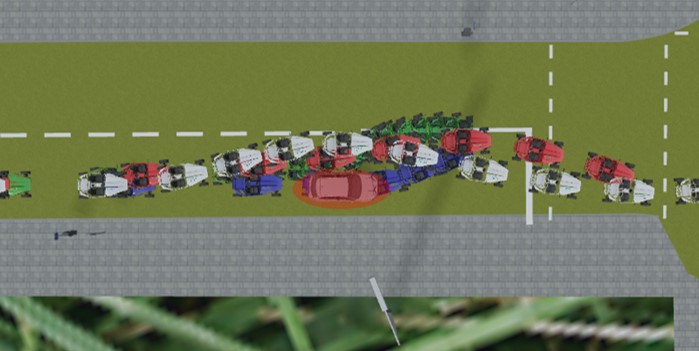}\label{fig:cbf_vs_pos_constraint_munich_b}}
	\caption{Closed-loop control with collision avoidance using CBF with RESAFE/COL (white) in comparison with standard position avoidance with RESAFE/COL (red), DMS (green) and PSC (blue): MPC-CBF ensures a natural double lane change. For a horizon length of $1.75\,\textrm{s}$ (Fig.~\ref{fig:cbf_vs_pos_constraint_munich_b}), only RESAEF/COL formulations avoid the obstacle safely}
	\label{fig:cbf_vs_pos_constraint_munich}
\end{figure}

\begin{figure}
	\begin{center}
%
%
\begin{tikzpicture}

\begin{axis}[%
width=.85\columnwidth,
height=3cm,
at={(0.0in,0.0in)},
scale only axis,
xmin=0.5,
xmax=4.5,
xtick={1,2,3,4},
xticklabels={{DMS},{PSC},{RESAFE/COL\\(ours) },{RESAFE/COL\\+CBF(ours)}},
xticklabel style = {font=\tiny,yshift=0.5ex, align=center},
ymin=-9.29225,
ymax=200,
ylabel near ticks,
ylabel style={font=\color{white!15!black}},
ylabel={Computation time [ms]},
extra y ticks={25, 50},
axis background/.style={fill=white},
xmajorgrids,
ymajorgrids
]
\addplot [color=black, dashed, forget plot]
  table[row sep=crcr]{%
1	98.519\\
1	162.041\\
};
\addplot [color=black, dashed, forget plot]
  table[row sep=crcr]{%
2	10.005\\
2	16.007\\
};
\addplot [color=black, dashed, forget plot]
  table[row sep=crcr]{%
3	7.9845\\
3	12.034\\
};
\addplot [color=black, dashed, forget plot]
  table[row sep=crcr]{%
4	10\\
4	14.04\\
};
\addplot [color=black, dashed, forget plot]
  table[row sep=crcr]{%
1	26.033\\
1	56.1085\\
};
\addplot [color=black, dashed, forget plot]
  table[row sep=crcr]{%
2	2.588\\
2	6\\
};
\addplot [color=black, dashed, forget plot]
  table[row sep=crcr]{%
3	1.963\\
3	5\\
};
\addplot [color=black, dashed, forget plot]
  table[row sep=crcr]{%
4	3.999\\
4	7\\
};
\addplot [color=black, forget plot]
  table[row sep=crcr]{%
0.875	162.041\\
1.125	162.041\\
};
\addplot [color=black, forget plot]
  table[row sep=crcr]{%
1.875	16.007\\
2.125	16.007\\
};
\addplot [color=black, forget plot]
  table[row sep=crcr]{%
2.875	12.034\\
3.125	12.034\\
};
\addplot [color=black, forget plot]
  table[row sep=crcr]{%
3.875	14.04\\
4.125	14.04\\
};
\addplot [color=black, forget plot]
  table[row sep=crcr]{%
0.875	26.033\\
1.125	26.033\\
};
\addplot [color=black, forget plot]
  table[row sep=crcr]{%
1.875	2.588\\
2.125	2.588\\
};
\addplot [color=black, forget plot]
  table[row sep=crcr]{%
2.875	1.963\\
3.125	1.963\\
};
\addplot [color=black, forget plot]
  table[row sep=crcr]{%
3.875	3.999\\
4.125	3.999\\
};
\addplot [color=blue, forget plot]
  table[row sep=crcr]{%
0.75	56.1085\\
0.75	98.519\\
1.25	98.519\\
1.25	56.1085\\
0.75	56.1085\\
};
\addplot [color=blue, forget plot]
  table[row sep=crcr]{%
1.75	6\\
1.75	10.005\\
2.25	10.005\\
2.25	6\\
1.75	6\\
};
\addplot [color=blue, forget plot]
  table[row sep=crcr]{%
2.75	5\\
2.75	7.9845\\
3.25	7.9845\\
3.25	5\\
2.75	5\\
};
\addplot [color=blue, forget plot]
  table[row sep=crcr]{%
3.75	7\\
3.75	10\\
4.25	10\\
4.25	7\\
3.75	7\\
};
\addplot [color=red, forget plot]
  table[row sep=crcr]{%
0.75	70.303\\
1.25	70.303\\
};
\addplot [color=red, forget plot]
  table[row sep=crcr]{%
1.75	7.004\\
2.25	7.004\\
};
\addplot [color=red, forget plot]
  table[row sep=crcr]{%
2.75	6\\
3.25	6\\
};
\addplot [color=red, forget plot]
  table[row sep=crcr]{%
3.75	7.999\\
4.25	7.999\\
};
\addplot [color=black, only marks, mark=+, mark options={solid, draw=red}, forget plot]
  table[row sep=crcr]{%
1	162.598\\
1	163.623\\
1	164.034\\
1	165.374\\
1	167.067\\
1	170.528\\
1	173.048\\
1	176.044\\
1	180.045\\
1	183.684\\
1	190.817\\
1	227.068\\
};
\addplot [color=black, only marks, mark=+, mark options={solid, draw=red}, forget plot]
  table[row sep=crcr]{%
2	16.036\\
2	16.521\\
2	16.964\\
2	16.995\\
2	16.996\\
2	16.997\\
2	16.997\\
2	17\\
2	17\\
2	17.001\\
2	17.007\\
2	17.009\\
2	17.009\\
2	17.014\\
2	17.037\\
2	17.358\\
2	17.513\\
2	17.964\\
2	17.992\\
2	17.998\\
2	17.999\\
2	17.999\\
2	18\\
2	18.001\\
2	18.002\\
2	18.002\\
2	18.003\\
2	18.003\\
2	18.004\\
2	18.006\\
2	18.034\\
2	18.044\\
2	18.135\\
2	18.512\\
2	18.997\\
2	18.998\\
2	19.005\\
2	19.011\\
2	19.268\\
2	19.518\\
2	19.521\\
2	19.968\\
2	20\\
2	20\\
2	20.002\\
2	20.549\\
2	20.974\\
2	20.993\\
2	20.998\\
2	21.964\\
2	21.996\\
2	21.999\\
2	22.039\\
2	22.553\\
2	23.003\\
2	23.009\\
2	23.553\\
2	23.996\\
2	24\\
2	24.515\\
2	24.994\\
2	24.998\\
2	25.995\\
2	26.004\\
2	26.034\\
2	26.15\\
2	26.547\\
2	27.001\\
2	28.513\\
2	28.998\\
2	34.001\\
};
\addplot [color=black, only marks, mark=+, mark options={solid, draw=red}, forget plot]
  table[row sep=crcr]{%
3	12.507\\
3	12.51\\
3	12.524\\
3	12.966\\
3	12.995\\
3	12.998\\
3	13\\
3	13\\
3	13\\
3	13.006\\
3	13.136\\
3	13.505\\
3	13.519\\
3	13.521\\
3	13.538\\
3	13.997\\
3	13.997\\
3	14.004\\
3	14.009\\
3	14.48\\
3	14.511\\
3	14.997\\
3	14.997\\
3	14.998\\
3	15.006\\
3	15.041\\
3	15.517\\
3	15.999\\
3	16.002\\
3	16.003\\
3	16.995\\
3	16.995\\
3	16.999\\
3	16.999\\
3	17\\
3	18.009\\
3	18.034\\
3	18.035\\
3	18.513\\
3	19\\
3	20.999\\
3	20.999\\
3	21.001\\
3	21.007\\
3	21.554\\
3	22.001\\
3	22.002\\
3	22.964\\
3	23.001\\
3	23.026\\
3	23.518\\
3	25.529\\
3	25.999\\
3	27.003\\
3	29\\
3	31.999\\
3	32\\
};
\addplot [color=black, only marks, mark=+, mark options={solid, draw=red}, forget plot]
  table[row sep=crcr]{%
4	14.508\\
4	14.511\\
4	14.512\\
4	14.996\\
4	14.998\\
4	14.999\\
4	14.999\\
4	14.999\\
4	14.999\\
4	15\\
4	15\\
4	15\\
4	15.001\\
4	15.001\\
4	15.001\\
4	15.001\\
4	15.001\\
4	15.002\\
4	15.002\\
4	15.002\\
4	15.004\\
4	15.004\\
4	15.006\\
4	15.007\\
4	15.007\\
4	15.035\\
4	15.035\\
4	15.04\\
4	15.261\\
4	15.507\\
4	15.51\\
4	15.512\\
4	15.625\\
4	15.995\\
4	15.997\\
4	15.998\\
4	15.999\\
4	15.999\\
4	15.999\\
4	16.001\\
4	16.001\\
4	16.001\\
4	16.002\\
4	16.003\\
4	16.006\\
4	16.114\\
4	16.582\\
4	16.997\\
4	16.998\\
4	17\\
4	17\\
4	17\\
4	17.004\\
4	17.005\\
4	17.009\\
4	17.034\\
4	17.511\\
4	17.999\\
4	18\\
4	18.003\\
4	18.008\\
4	18.513\\
4	18.521\\
4	19.009\\
4	19.481\\
4	20\\
4	20.512\\
4	20.519\\
4	20.957\\
4	21\\
4	21.001\\
4	21.006\\
4	21.006\\
4	21.008\\
4	22.52\\
4	23.286\\
4	23.971\\
4	24.008\\
4	24.51\\
4	24.51\\
4	24.997\\
4	25.515\\
4	25.998\\
4	26\\
4	27.998\\
4	28.007\\
4	29\\
4	36.002\\
};
\addplot [color=black, dashed]
  table[row sep=crcr]{%
0	50\\
1	50\\
2	50\\
3	50\\
4	50\\
5	50\\
};
\addlegendentry{Sampling period}
\end{axis}

\begin{axis}[%
width=.7\columnwidth,
height=3cm,
at={(0in,0in)},
scale only axis,
xmin=0,
xmax=1,
ymin=0,
ymax=1,
axis line style={draw=none},
ticks=none,
axis x line*=bottom,
axis y line*=left
]
\end{axis}
\end{tikzpicture}%
		\caption{Comparison of the different numerical methods and collision avoidance approaches with a long horizon length, over all instances of the closed-loop scenarios} 
		\label{fig:computation_time_comparison}
	\end{center}
\end{figure}
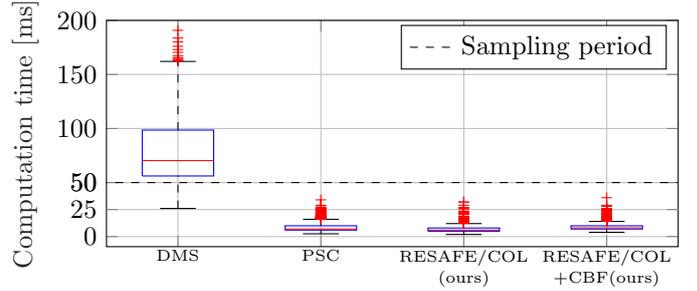

The closed-loop performance of the different controllers is shown in Figure~\ref{fig:cbf_vs_pos_constraint_munich} in which the formulations with RESAFE/COL outperform DMS and PSC, specifically in the event of a safety-critical situation with a short look-ahead of the MPC. 
Moreover, the developed RESAFE/COL approach is intended for real-time platforms. Therefore, we first assess the required computational effort in comparison with DMS and PSC. For this, we run a multitude of scenarios, to mimic an urban driving style: path following at $20\,\textrm{m/s}$, intersections, roundabouts, and a multitude of road users such as dynamic vehicles, pedestrians, and dangerous objects to test the safety-critical aspect.
For a considerably large horizon length $t_f = 3\,\textrm{s}$, the NMPC is tasked to solve in a receding horizon at the sampling period of $50 \, \textrm{ms}$. In the boxplot of the computation time in Figure~\ref{fig:computation_time_comparison}, the collocation methods have a clear computational advantage over DMS. Albeit solving to convergence, the PSC method does not satisfy the continuous path constraints fully, and therefore does not reflect the original continuous time OCP, ending up crashing with the obstacle at high-speed. Using RESAFE/COL on the other hand, both open-loop trajectories computed at every instance, and the closed-loop performance, satisfy the collision constraints described in the continuous time OCP. Using only the barrier function~\eqref{eq:barrier_constraint} as a constraint, RESAFE/COL with a mean of computation time at $7\, \textrm{ms}$ is 7 times faster than DMS, and all the computation instances are well below the sampling period of $50\, \textrm{ms}$. Similarly, by including the CBF constraint~\eqref{eq:cbf_constraint}, despite an increase in the computation time to $11\,\textrm{ms}$. It is worth nothing that the computation time using RESAFE/COL is not highly sensitive to the change of active set during the appearance of obstacles. This is reflected by the small spread in computation time in comparison to the large spread using DMS.  

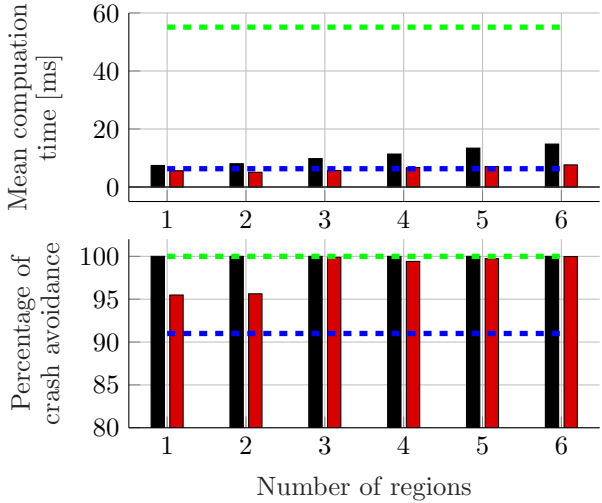
\begin{figure}
	\begin{center}
%
%
\definecolor{mycolor1}{rgb}{0.00000,0.0,0.0}%
\definecolor{mycolor2}{rgb}{0.85000,0.0,0.0}%
\begin{tikzpicture}

\begin{axis}[%
width=0.7\columnwidth,
height=2.5cm,
at={(0.0in,3.0cm)},
scale only axis,
bar shift auto,
xmin=0.514285714285714,
xmax=6.48571428571429,
ymin=-5,
ymax=60,
ylabel style={font=\color{white!15!black}, align=center},
ylabel={Mean compuation \\ time [ms]},
axis background/.style={fill=white},
axis x line*=bottom,
axis y line*=left,
xmajorgrids,
ymajorgrids,
legend style={at={(-0.2,0.941)}, anchor=south west, legend columns=4, legend cell align=left, align=left, draw=white!15!black}
]
\addplot[ybar, bar width=5, fill=mycolor1, draw=black, area legend] table[row sep=crcr] {%
1	7.43190174854288\\
2	8.02832472939217\\
3	9.78902747710242\\
4	11.3663555370525\\
5	13.4064837635304\\
6	14.8037452123231\\
};
\addplot[forget plot, color=white!15!black] table[row sep=crcr] {%
0.514285714285714	0\\
6.48571428571429	0\\
};

\addplot[ybar, bar width=5, fill=mycolor2, draw=black, area legend] table[row sep=crcr] {%
1	5.60588676103247\\
2	5.07839633638635\\
3	5.68805661948376\\
4	6.69946877601998\\
5	7.05719900083264\\
6	7.62098251457119\\
};
\addplot[forget plot, color=white!15!black] table[row sep=crcr] {%
0.514285714285714	0\\
6.48571428571429	0\\
};

\addplot [color=blue, dashed, line width=2.0pt]
  table[row sep=crcr]{%
1	6.30155648071052\\
2	6.30155648071052\\
3	6.30155648071052\\
4	6.30155648071052\\
5	6.30155648071052\\
6	6.30155648071052\\
};

\addplot [color=green, dashed, line width=2.0pt]
  table[row sep=crcr]{%
1	55.1070933943936\\
2	55.1070933943936\\
3	55.1070933943936\\
4	55.1070933943936\\
5	55.1070933943936\\
6	55.1070933943936\\
};

\end{axis}

\begin{axis}[%
width=0.7\columnwidth,
height=2.5cm,
at={(0.0in,0.0in)},
scale only axis,
bar shift auto,
xmin=0.514285714285714,
xmax=6.48571428571429,
xlabel style={font=\color{white!15!black}},
xlabel={Number of regions},
ymin=80,
ymax=102,
ylabel style={font=\color{white!15!black}, align=center},
ylabel={Percentage of \\ crash avoidance},
axis background/.style={fill=white},
axis x line*=bottom,
axis y line*=left,
xmajorgrids,
ymajorgrids
]
\addplot[ybar, bar width=5, fill=mycolor1, draw=black, area legend] table[row sep=crcr] {%
1	100\\
2	100\\
3	100\\
4	100\\
5	100\\
6	100\\
};
\addplot[forget plot, color=white!15!black] table[row sep=crcr] {%
0.514285714285714	0\\
6.48571428571429	0\\
};
\addplot[ybar, bar width=5, fill=mycolor2, draw=black, area legend] table[row sep=crcr] {%
1	95.49\\
2	95.63\\
3	99.886\\
4	99.41\\
5	99.73\\
6	99.972\\
};
\addplot[forget plot, color=white!15!black] table[row sep=crcr] {%
0.514285714285714	0\\
6.48571428571429	0\\
};
\addplot [color=green, dashed, line width=2.0pt, forget plot]
  table[row sep=crcr]{%
1	100\\
2	100\\
3	100\\
4	100\\
5	100\\
6	100\\
};
\addplot [color=blue, dashed, line width=2.0pt, forget plot]
  table[row sep=crcr]{%
1	91\\
2	91\\
3	91\\
4	91\\
5	91\\
6	91\\
};
\end{axis}
\end{tikzpicture}%
		\caption{Sensitivity of computation time and closed-loop performance in terms of collision avoidance to the number of regions in RESAFE/COL, with a large horizon length: RESAFE/COL with CBF (black); RESAFE/COL with position constraint only (red), PSC order 5 (blue) and DMS with 60 nodes (green)} 
		\label{fig:computation_time_comparison_shorthorizon}
	\end{center}
\end{figure}

In addition, we perform a sensitivity analysis over the number of regions utilized for extrema estimation in Theorem~\ref{thm:resafe_col} and \eqref{eq:resafecol_nlp}, with the results shown in Figure~\ref{fig:computation_time_comparison_shorthorizon}. The percentage of crash avoidance is calculated as the ratio between the total time the vehicle is outside the enlarged obstacles' ellipse over the total duration in which the vehicle detects an obstacle closer than 30 meters away. For the percentage of crash avoidance, we consider the closed-loop performance of the vehicle. The NMPC for the vehicles with DMS (green) and PSC (blue) remain unchanged with 60 shooting nodes, and polynomial of order 5 respectively. As the OCP with the CBF constraint is more complex than the position avoidance counterpart, the computation time is more sensitive to the number of regions. That is due to tightening on the extrema of both the barrier constraint and CBF constraints, which activates the NLP constraints. However, computation time remains well below the sampling period of $50\,\textrm{ms}$ confirming the real-time feasibility of the complex OCP, and a performance improvement between 5 and 7 times in terms of computation time in comparison with DMS. 
As the formulation through a TLS and the exploitation of the convex-hull are intended to not deteriorate the control performance, we also visualize the percentage of crash avoidance in Figure~\ref{fig:computation_time_comparison_shorthorizon}. With a dense sampling of 60 nodes, DMS is able to maintain a 100\% crash avoidance in closed-loop, still at the risk of constraint violation between the shooting nodes in the open-loop prediction. As from Proposition \ref{prop:decreasing_error_resafecol}, the extrema approximation with RESAFE/COL decreases as the number of regions increases, the percentage of crash avoidance with RESAFE/COL increases, and reaches 99.5\% starting 3 regions. That is, the NMPC solves for more dynamic trajectories near the limit of handling, with less conservatism as the number of regions increases. Finally, with PSC, the percentage of crash avoidance is at 91\% with noticeable continuous time OCP constraints' violations in the open-loop trajectories as the constraints are only satisfied at 6 collocation points. 

\subsection{Comparison between position and CBF constraints}
\begin{figure}
	\begin{center}
		\input{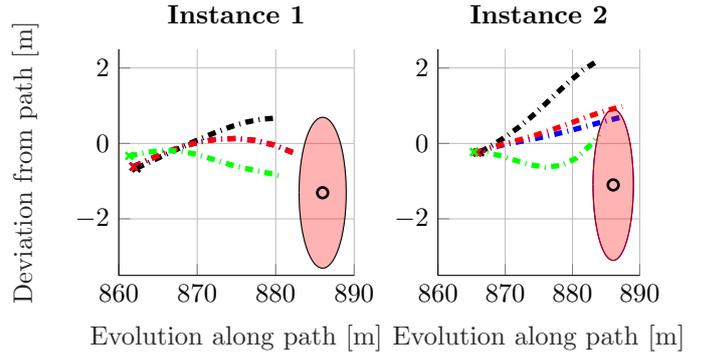}    
		\caption{Open-loop trajectory during collision avoidance using forward invariant CBF with RESAFE/COL (black) in comparison with position constraint with RESAE/COL (red), PSC(blue) and DMS (green)} 
		\label{fig:ol_trajectory_cbf}
	\end{center}
\end{figure}

For the vehicles with the position (barrier) constraint \eqref{eq:barrier_constraint} only, Vehicle 3 (red) initiates the avoidance maneuver earlier than Vehicles 2 and 4 as seen in Figure~\ref{fig:cbf_vs_pos_constraint_munich_a}, and has a safer escape than Vehicles 2 and 4 that pass tightly along the obstacles' ellipse. The inclusion of the CBF constraint does not considerably change the closed-loop performance of the system in terms of collision avoidance. This is due the long horizon length, allowing the vehicle to react early to upcoming obstacles. Among the approaches with RESAFE/COL, Vehicle 3 exhibits a larger deviation from the path and a longer recovery time after crossing the parked vehicle. The applied control effort are therefore slightly larger and more aggressive than with the CBF counterpart. 
In order to verify the benefit of the CBFs constraints in terms of satisfying the collision avoidance in the infinite horizon ahead, we reduce the horizon length from $t_f=3\,\textrm{s}$ to $t_f=1.75\,\textrm{s}$. That is, at full speed of $20\,\textrm{m/s}$, the NMPC can only predict for a distance of $35\,\textrm{m}$ ahead.
In Figure \ref{fig:cbf_vs_pos_constraint_munich_b}, the closed-loop performance of all 4 vehicles is visualized, in a safety-critical situation around a parked vehicle highlighted in red. The first two direct realizations are that PSC and DMS fail completely to avoid the obstacle. That is, even when the open-loop trajectories are safe, the model mismatch between the NMPC prediction model and the Digital Twin, causes both vehicles to overestimate their dynamic abilities. Once close to the obstacle, both vehicles engage in an emergency brake maneuver. In the case of PSC, and given that the collision avoidance constraint is satisfied at finite points, the NMPC barely commands the vehicle to steer away from the obstacle, causing a rear collision. Unlike the NMPC formulated with RESAFE/COL (red, white) that steer the vehicle away on time.
For the two NMPCs formulated with RESAFE/COL, the controller with the CBF constraint (white), and for the same enlarged size of the vehicle, is more cautious and engages in a slow-down, avoid, and recover maneuver that resembles the ISO 3888-1 standard double lane change. The controller with only the position constraint avoids the obstacle, but just tightly at the limits of the ellipse, without decreasing the speed of the vehicle. The open-loop prediction at two instances near the obstacle are shown in Figure~\ref{fig:ol_trajectory_cbf}: the CBF constraint ensures a forward invariance of the predicted trajectory beyond the NMPC horizon, allowing a smoother and earlier deviation from the obstacle.

\begin{figure}
	\centering
	\subfloat[][Position constraint]{	\includegraphics[width=0.22\textwidth]{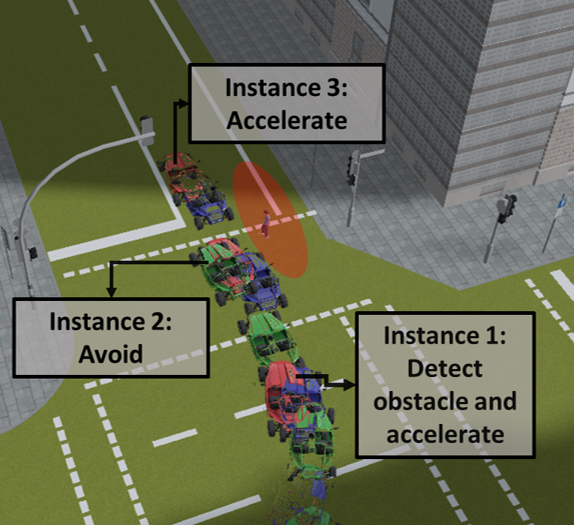}}
	\subfloat[][CBF constraint]{	\includegraphics[width=0.22\textwidth]{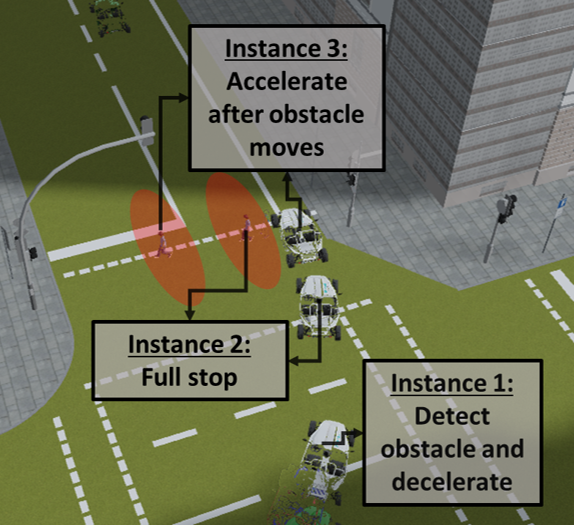}}
	\caption{Collision avoidance at an intersection with and without CBF. The CBF formulation embeds both motion planning and control as it performs a stop-and-go}
	\label{fig:cbf_nocbf_intersection}
\end{figure}

Furthermore, we test the capabilities of the controller at an intersection. Following a highly curved turn, an obstacle suddenly appears in front of all vehicles as depicted in Figure \ref{fig:cbf_nocbf_intersection}. The ability of the CBF formulation to incorporate mid-level motion planning by altering the velocity profile, and low-level control tracking is put to highlight. With the standard collision avoidance constraint, all three vehicles avoid the obstacle through high steering and continuous acceleration. Albeit being a successful collision avoidance, the evasive maneuver is uncomfortable and unnatural. One could follow a hierarchical framework with a planner for velocity decision-making, however, this generally comes in the form of a non-scalable rule-based or heuristic module.  Alternatively, with the CBF constraint, the vehicle performs a stop-and-go maneuver, by slowing down to the pedestrian, coming to a natural and smooth full stop without an emergency brake, and then accelerating to recover the velocity and path tracking as the obstacle moves. This distinction between velocity and path tracking allowing such a decision-making is possible through the path following formulation that plans over space rather than time as in Figure~\ref{fig:curvilinear_frame}.

\captionof{table}{NMPC's performance during safety-critical maneuvers with a short prediction horizon of $1.75\,\textrm{s}$, at $20 \, \textrm{m/s}$\label{tab:shorthorizon_benchmark}}
\begin{table}[hb!] 
	\begin{center}
		\begin{tabular}{|c||c|c|}
			\hline
			Method & \makecell{Mean computation \\ time [ms]} & \makecell{Percentage of \\crash}\\\hline
			DMS & 37.6 & 14.29  \\\hline
			PSC & 7.8 & 4.09  \\ \hline
			RESAFE/COL & 7.7 & 1.28  \\\hline
			\thead{RESAFE/COL+CBF} & \thead{12.3} & \thead{0}  \\\hline
		\end{tabular}
	\end{center}
\end{table}

Finally, we showcase the computational and control performance advantages offered by RESAFE/COL in a safety-critical situation. From Table~\ref{tab:shorthorizon_benchmark}, RESAFE/COL is approximately 5 times faster than DMS for the same formulation, and 3 time faster for the more complex formulation with CBF. The computational benefit is also complemented by the success of collision avoidance: while it is at 100\% with RESAFE/COL+CBF similar to the case of long horizon, RESAFE/COL with position constraint is able to decrease the total percentage of in-crash instances by 91\% from 14.29\% to 1.28\%.

\section{Conclusion}\label{sec:conclusion}
A safety-critical model predictive controller based on control barrier functions has been proposed in this paper to tackle urban autonomous vehicle driving in the presence of road users. We presented RESAFE/COL a novel collocation based transcription approach allowing to satisfy the nonlinear constraints of the continuous time OCP over the full prediction horizon. We validate the approach in high-fidelity vehicle and environment simulators. The navigation of the vehicle relies on a special set of splines that can guarantee a safe tracked trajectory in the look-ahead of the NMPC, with spatio-temporal predictions in the error frame. The proposed transcription approach makes the problem more tractable for real-time applications without decreasing the performance, as it is shown to outperform the direct multiple shooting method in terms of computation time and delivers a safer and more stable navigation than standard collision avoidance formulations that suffer from the shortness of the NMPC horizon.


\bibliography{ifacconf}             

\end{document}